\documentclass[nolinenumbers]{aastex631}

\begin{document}

\title{Sesquinary Catastrophe For Close-In Moons with Dynamically Excited Orbits}

\correspondingauthor{Matija \'Cuk}
\email{mcuk@seti.org}

\author[0000-0003-1226-7960]{Matija \'Cuk}
\affiliation{SETI Institute \\
189 N Bernardo Ave\\
Mountain View, CA 94043, USA}

\author{Douglas P. Hamilton}
\affiliation{Department of Astronomy\\
University of Maryland\\ 
Physical Sciences Complex\\ 
College Park MD 20742, USA}

\author{David A. Minton}
\affiliation{Department of Earth, Atmospheric and Planetary Sciences \\
Purdue University  \\
550 Stadium Mall Drive\\ 
West Lafayette, IN 47907, USA}

\author{Sarah T. Stewart}
\affiliation{Department of Earth and Planetary Sciences\\
University of California Davis\\
One Shields Avenue\\
Davis, CA 95616, USA}



\begin{abstract}
We identify a new mechanism that can lead to the destruction of small, close-in planetary satellites. If a small moon close to the planet has a sizable eccentricity and inclination, its ejecta that escape to planetocentric orbit would often re-impact with much higher velocity due to the satellite's and the fragment's orbits precessing out of alignment. If the impacts of returning ejecta result in net erosion, a runaway process can occur which may end in disruption of the satellite, and we term this process ``sesquinary catastrophe''. We expect the moon to re-accrete, but on an orbit with significantly lower eccentricity and inclination. We find that the large majority of small close-in moons in the Solar System, have orbits that are immune to sesquinary catastrophe. The exceptions include a number of resonant moonlets of Saturn for which resonances may affect the velocities of  re-impact of their own debris. Additionally, we find that Neptune's moon Naiad (and to a lesser degree, Jupiter's Thebe) must have substantial internal strength, in line with prior estimates based on Roche limit stability. We also find that sesquinary instability puts important constraints on the plausible past orbits of Phobos and Deimos or their progenitors. 
\end{abstract}



\section{Introduction} \label{sec:intro}

Regular satellites of the giant planets are the moons that have near-circular orbits that are close to the plane of the planet's equator. They are distinguished from irregular satellites which orbit on distant, usually highly eccentric and inclined orbits; irregular satellites are generally thought to be captured from heliocentric orbits \citep{nic08}. Among regular satellites there is a distinct class of so-called {\it ring-moons} which are typically close to the planet (within 3-4 planetary radii), small (with radii under 200~km), have irregular shapes, and sometimes interact with planetary rings. Arguably, the Martian moons Phobos and Deimos have many properties of ring moons, despite Deimos orbiting beyond 6 Mars radii. As Mars is more dense and a slower rotator than the giant planets, both its Roche limit (which is the closest orbital distance possible for strengthless bodies) and the synchronous orbit are further away from the planet than they are for the gas giant, with the Martian synchronous orbit being just interior to that of Deimos.

Ring moons are unique astronomical objects due to the combination of three factors: 1) they are small and have weak gravity, 2) their orbital velocities are large due to orbiting close to major planets, and 3) their sizes are not negligible compared to their orbital distances, leading to relatively quick collisions between objects on crossing orbits. Other classes of Solar System objects may share two but not all three of these characteristics. For example, major satellites lack the first property as their escape velocities are large compared with those of ring moons. Asteroidal moons lack the second property as they often orbit at speeds below 1 m/s \citep{wal15}, and the asteroids orbiting the Sun lack the third property, as the probability of collision between any given pair of small bodies on heliocentric crossing orbits in extremely small. The irregular satellites of the giant planets are more similar to asteroids than inner moons for our purposes here. Each giant planet has numerous irregulars on crossing orbits, likely undergoing collisional evolution through mutual impacts \citep{bot10}. Collisions between prograde and retrograde irregular satellites are likely to be particularly energetic, dominating their collisional evolution.

This physical environment of ring moons makes them susceptible to sesquinary cratering, first proposed by \citet{zah08}. Craters formed by impactors unrelated to the target are considered primary, while craters made by sub-orbital ejecta launched by a primary impact are termed secondary \citep{rob64}. Sesquinary\footnote{Term derived from Latin for ``one and a half'', \citet{zah08}.} craters are those caused by ejecta from the target satellite that has escaped its gravitational pull but has reimpacted after orbiting the parent planet on an independent orbit. It is possible that many small craters on Saturn's moon Mimas (which is larger and more distant than a typical ring-moon) are sesquinaries \citep{alv05, bie12, alv17}, and this is also likely the case for many ring-moons.

The orbits of ring moons are close to being circular and planar, but they are not perfectly so. Some ring moons are in orbital resonances with each other, notably Neptune's Naiad and Thalassa \citep{bro20} and Uranus's Belinda and Perdita \citep{fre17, cuk22}. Others may have been affected in the past by resonances with other moons \citep{ban92, ham01, zha07, zha08} or, in case of Mars, resonances with the Sun and Martian rotation \citep{yod82}. In any case, eccentricities and inclinations (in radians) on the order of a percent are often found among rings moons\footnote{Inclinations are assumed to be measured relative to local Laplace plane, i.e. the plane around which a moon's orbital plane precesses; for most inner satellites their Laplace plane is close, but not identical, to their parent planet's equatorial plane}. While relatively small, these deviations from dynamically cold orbits can dominate the re-impact velocity of sesquinary impactors. Ejecta from an an inclined and/or eccentric ring-moon would largely inherit the moon's orbital parameters, as the ejection velocity is assumed to be on the order of the satellite's escape velocity. As the orbits of the moon and the fragment precess out of alignment, the re-impact velocity will be determined by the inclination and/or eccentricity of the moon's orbit. If the reimpact happens at a velocity that is above the threshold for erosion \citep[as opposed to partial accretion, see][]{lei12, ste12}, the moon will gradually lose more material in a feedback loop. The end outcome is not clear, but may even include disruption and re-accretion of the moon. We term this theoretical phenomenon ``sesquinary catastrophe'' and we will explore its physics and implication for the Solar System satellites in the following Sections.

\section{Sesquinary Catastrophe In the Present Solar System}

\begin{figure}[h!]
\vspace{-.15truein}
\hspace{0truein}
\begin{minipage}{2.8truein}
\centering
\includegraphics[scale=.8]{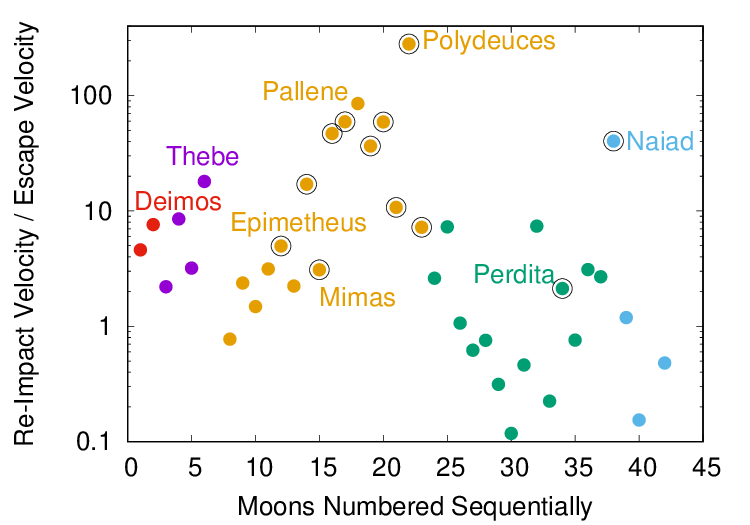}
\end{minipage}
\hspace{1.1 truein}
\vspace{-3.5in}
\begin{minipage}{2.5truein}
\centering
\vspace{-.1in}
\caption{The approximate velocity of ejecta returning to the moon as sesquinary impactors normalized to the escape velocity ($q_v$, Eq.~\ref{eq:q}) for each of the small inner moons as numbered in Table \ref{moons}. Point colors indicate the parent planet: Mars (red), Jupiter (purple), Saturn (orange), Uranus (green) and Neptune (cyan). Black circles indicate satellites in a resonance with a more massive moon. Only the moons with $q_v>0.1$ are included in this plot.}\label{plan}
\end{minipage}
\vspace{3.4in}
\end{figure}

In order to assess the potential relevance of the sesquinary catastrophe mechanism, we can compare the orbital and physical parameters of the small inner moons of the planets with the expected threshold for the sesquinary catastrophe. Figure \ref{plan} plots the ratio of typical ejecta re-impact velocity to escape velocity for Phobos, Deimos and a number of small inner moons of the giant planets. The value on the y-axis is simply calculated as
\begin{equation}
    q_v=\sqrt{e^2+\sin^2{i}}(v_{orb} / v_{esc})
    \label{eq:q}
\end{equation} 
for each moon. The numbering of the satellites as well as $q_v$ values are listed in Table \ref{moons}. 

\begin{figure}[h!]
\vspace{-.15truein}
\hspace{0truein}
\begin{minipage}{2.8truein}
\centering
\includegraphics[scale=.8]{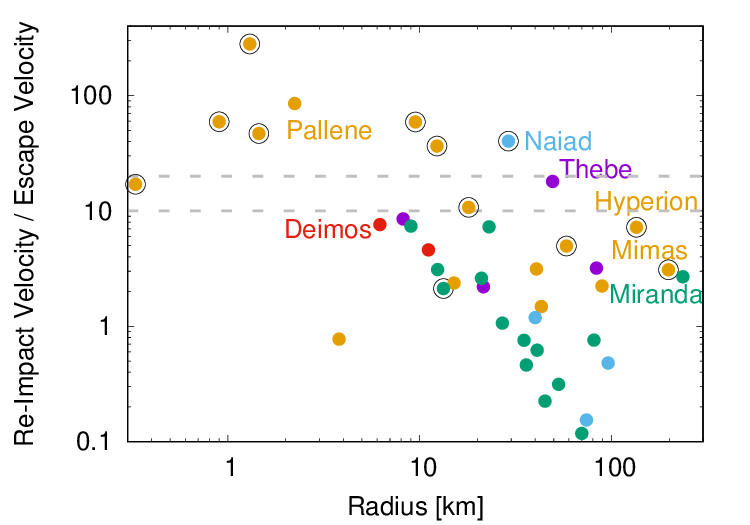}
\end{minipage}
\hspace{1.1 truein}
\vspace{-3.5in}
\begin{minipage}{2.5truein}
\centering
\vspace{-.1in}
\caption{The approximate velocity of ejecta returning to the moon as sesquinary impactors normalized to the escape velocity ($q_v$) versus the radius for our sample of small, close-in planetary satellites. Point colors indicate the parent planet: Mars (red), Jupiter (purple), Saturn (orange), Uranus (green) and Neptune (cyan). Black circles indicate satellites in a resonance with a more massive moon. Only the moons with $q_v>0.1$ are included in this plot.}\label{size}
\end{minipage}
\vspace{3.4in}
\end{figure}

Fig.~\ref{plan} shows that the large majority of small inner moons have the ratio $q_v$ below about 10-20. $q_v=10$ is approximately what is expected as threshold for net erosion when the impactor is much smaller than  the target \citep{ste12}. While most of the moons plotted in Fig. \ref{plan} are small in size, there is almost three orders of magnitude range in radius between Uranus's Miranda and Saturn's Aegaeon. Are the high-$q_v$ moons mostly small moons with low escape velocities? In order to test this possibility, we examined the parameter $q_v$ against the moon's radius in Fig. \ref{size}. 

Figure \ref{size} shows that there is a clear anti-correlation between a moon's parameter $q_v$ and its radius. This is expected, as $q_v$ is inversely proportional to the moon's escape velocity. However, not all moons with $q_v>10$ are small. Jupiter's Thebe and Neptune's Naiad are above the median size for our sample but are still potentially vulnerable to the sesquinary catastrophe due to their excited orbits.
 
\begin{figure}[h!]
\vspace{-.15truein}
\hspace{0truein}
\begin{minipage}{2.8truein}
\centering
\includegraphics[scale=.8]{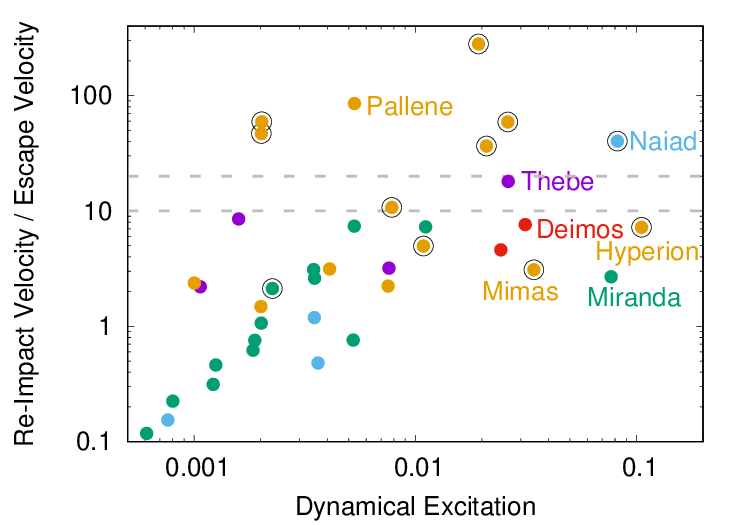}
\end{minipage}
\hspace{1.1 truein}
\vspace{-3.5in}
\begin{minipage}{2.5truein}
\centering
\vspace{-.1in}
\caption{The approximate velocity of ejecta returning to the moon as sesquinary impactors normalized to the escape velocity ($q_v$) versus dynamical excitation ($\eta=\sqrt{e^2+\sin^2i}$) for the same set of small, close-in moons as in Fig. \ref{plan}. Point colors indicate the parent planet: Mars (red), Jupiter (purple), Saturn (orange), Uranus (green) and Neptune (cyan). Black circles indicate satellites in a resonance with a more massive moon. Only the moons with $q_v>0.1$ are included in this plot.}\label{amd}
\end{minipage}
\vspace{3.4in}
\end{figure}

 Most of the moons with $q_v>10$ are in resonance with a larger satellite. These include a number of small moons in the Saturnian system, as well as Neptune's innermost known moon Naiad. There are different ways resonances can affect the potential sesquinary catastrophe and affect the value or applicability of the parameter $q_v$. One is that resonant moons may simply have higher eccentricities and inclinations, which would make the resonant moons over-represented among high-$q_v$ objects. To evaluate this possibility, we considered the $q_v$ parameters against the total amount of dynamical excitation measured by the quantity $\eta=\sqrt{e^2+\sin^2{i}}$ in Fig. \ref{amd}. 
While some of the dynamically most excited moons such as Hyperion and Naiad are indeed resonant, dynamical excitation alone cannot explain the overabundance of resonant moons among high-$q_v$ objects. Resonant moons cover a wide range of dynamical excitation, and some of the high-$q_v$ moons are not on particularly excited orbits. In particular, small moons of Saturn that orbit between the major moons (Pallene, Methone, Anthe, Aegaeon, Telesto, Calypso, Helene, and Polydeceus) all have high $q_v$ values, and are all resonant with the notable exception of Pallene, but mostly do not have exceptionally large $e$ or $i$.  

\begin{figure}[h!]
\vspace{-.15truein}
\hspace{0truein}
\begin{minipage}{2.8truein}
\centering
\includegraphics[scale=.8]{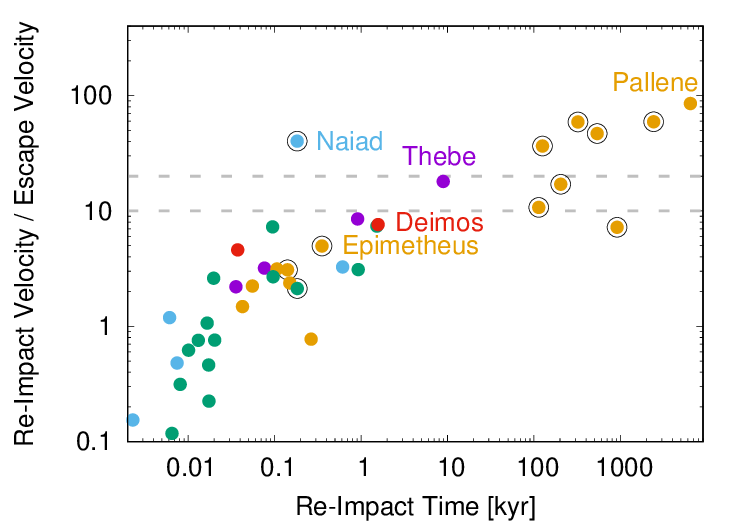}
\end{minipage}
\hspace{1.1 truein}
\vspace{-3.5in}
\begin{minipage}{2.5truein}
\centering
\vspace{-.1in}
\caption{The approximate velocity of ejecta returning to the moon as sesquinary impactors normalized to the escape velocity ($q_v$) versus the timescale for re-impact ($\tau$, Eq.~\ref{eq:tau}) for small, close-in planetary satellites. Point colors indicate the parent planet: Mars (red), Jupiter (purple), Saturn (orange), Uranus (green) and Neptune (cyan). Black circles indicate satellites in a resonance with a more massive moon. Only the moons with $q_v>0.1$ are included in this plot.}\label{sesq}
\end{minipage}
\vspace{3.4in}
\end{figure}

Another way a resonance can affect the vulnerability of moons to sesquinary catastrophe is through dynamics of ejecta. Material coming off the smaller moon may still be in the resonance, and the perturbations of the larger resonant companion may dominate the orbit of the ejecta and keep it from drifting far from that of the source moon. In other words, for many of the resonant moons the forced eccentricity may be larger than the free one, preventing the misalignment of pericenters of the ejecta and the moon. This is likely the case for Saturnian small moons that populate the right-hand side of the plot that are either Trojans or are in corotation resonances with Mimas. For example, \citep{cal21} find that Methone's orbit is aligned with that of Mimas due to substantial forced eccentricity. However, this promising explanation for the possible immunity of Saturn's small resonant moons to sesquinary catastrophe cannot be applied to Pallene, which is in may ways similar to its resonant siblings Anthe and Methone, but is not currently in the resonance and its eccentricity is largelly free. Why is Pallene not rapidly destroyed by its own ejecta? It is natural to hypothesize that Pallene may have formed in a resonance and then somehow escaped it. How much time would it be required for sesquinary catastrophe to gather pace? In Fig. \ref{sesq} we compare the parameter $q_v$ for each moon in our sample to a rough estimate of the time required for ejecta to reimpact. 

The collisional timescale in Fig. \ref{sesq} is calculated as 
\begin{equation}
\tau= (P /\Delta) \delta e \delta i (a/R)^2,    
\label{eq:tau}
\end{equation}
where $P$ is orbital period, $\Delta=v_{esc}/v_{orb}$ quantifies difference in semi-major axes of the moon and ejecta, $a \delta e= a \sqrt{\Delta^2+e^2}$ and $a \delta i = a \sqrt{\Delta^2+\sin^2{i}}$ are sides of the ``box''the ejecta can be in during conjunction, and $R$ is the radius of the moon (factors of order unity have been ignored in this approximation for $\tau$). Polydeuces is beyond the right edge of the plot (Table \ref{moons}).

\begin{deluxetable*}{rlccrrccrrr}
\tablenum{1}
\vspace{-.2in}
\tablecaption{Orbital and physical data for the a selection of planetary satellites, as well as derived quantities plotted in Figures \ref{plan}-\ref{sesq}. All orbital elements, as well as radii and masses of unlabelled moons are obtained from the Jet Propulsion Laboratory's Solar System Dynamics website on March 10, 2021. Radii and masses of the moons with asterisks are from \citet{tho20}. Mean radii of Uranian inner moons (with daggers) are the same as in \citet{cuk22}, originally from \citet{sho06} and \citet{kar01}, and the corresponding masses assume $\rho=860$~kg~m$^{-3}$ \citep{cha17}. Anthe's radius, $e$ and $i$ are from \citet{coo08} and its mass assumes $\rho=500$~kg~m$^{-3}$.Other sources of orbital elements are indicated by superscript: P -- \citet{por03}, J -- \citet{jac08},  H -- \citet{hed10}, J98 -- \citet{jac98}, F -- \citet{fre17}, B -- \citet{bro20}.\label{moons}}
\tablewidth{0pt}
\tablehead{\colhead{} & \colhead{Moon} & \colhead{Planet} & \colhead{Radius} & \colhead{GM} &\colhead{Semimajor} & \colhead{Eccentricity} & \colhead{Inclination} & \colhead{$q_v$} & \colhead{$\eta$} & \colhead{$\tau$}\\
\colhead{} & \colhead{Name} & \colhead{} & \colhead{[km]} & \colhead{[km$^{3}$~s$^{-2}$]} &\colhead{Axis $a$ [km]} &  \colhead{$e$} & \colhead{i [$^{\circ}$]} & \colhead{} & \colhead{} & \colhead{[yr]}}
\startdata 
1 & Phobos     & M &  11.1 & 0.710E-3 &    9400 & 0.015 & 1.1 &   4.6 & 0.024 & 0.38E+2 \\
2 & Deimos     & M &   6.2 & 0.960E-4 &   23500 & 0.000 & 1.8 &   7.6 & 0.031 & 0.16E+4 \\
3 & Metis      & J &  21.5 & 0.250E-2 &  128000 & 2E-4$^{\rm P}$ & 0.06$^{\rm P}$ &   2.2 & 0.000 & 0.35E+2 \\
4 & Adrastea   & J &   8.2 & 0.140E-3 &  129000 & 0.001$^{\rm P}$ & 0.03$^{\rm P}$ &   8.5 & 0.002 & 0.91E+3 \\
5 & Amalthea   & J &  83.5 & 0.165E+0 &  181400 & 0.003 & 0.4 &   3.2 & 0.008 & 0.76E+2 \\
6 & Thebe      & J &  49.3 & 0.300E-1 &  221900 & 0.018 & 1.1 &  18.0 & 0.026 & 0.89E+4 \\
7 & Pan        & S &  13.7 & 0.290E-3 &  133600 & 1E-5$^{\rm J}$ & 1E-4$^{\rm J}$ &   0.0 & 0.000 & 0.58E+2 \\
8 & Daphnis$^*$   & S &   3.8 & 0.510E-5 &  136500 & 3E-5$^{\rm J}$ & 4E-3$^{\rm J}$ &   0.8 & 0.000 & 0.26E+3 \\
9 & Atlas      & S &  15.1 & 0.370E-3 &  137700 & 0.001 & 3E-3$^{\rm J}$ &   2.4 & 0.001 & 0.15E+3 \\
10 & Prometeus  & S &  43.1 & 0.107E-1 &  139400 & 0.002 & 8E-3$^{\rm J}$ &   1.5 & 0.002 & 0.43E+2 \\
11 & Pandora    & S &  40.6 & 0.926E-2 &  141700 & 0.004 & 0.05$^{\rm J}$ &   3.1 & 0.004 & 1.07E+2 \\
12 & Epimetheus & S &  58.2 & 0.350E-1 &  151400 & 0.009 & 0.35$^{\rm J}$&   5.0 & 0.011 & 0.35E+3 \\
13 & Janus      & S &  89.2 & 0.127E+0 &  151500 & 0.007 & 0.16$^{\rm J}$ &   2.3 & 0.008 & 0.55E+2 \\
14 & Aegeaeon$^*$  & S &   0.33 & 0.520E-8 &  167500 & 2E-4$^{\rm H}$ & 1E-3$^{\rm H}$ &   17.0 & 0.000 & 0.20E+6 \\
15 & Mimas      & S & 198. & 0.250E+1 &  186000 & 0.020 & 1.60 &   3.1 & 0.034 & 0.14E+3 \\
16 & Methone$^*$   & S &   1.45 & 0.260E-6 &  194700 & 0.002 & 0.01$^{\rm J}$ &  46.6 & 0.002 & 0.54E+6 \\
17 & Anthe$^{\ddagger}$  & S &   0.9 & 0.100E-6 &  198100 & 0.002 & 0.02 &  58.7 & 0.002 & 0.24E+6 \\
18 & Pallene$^*$  & S &   2.23 & 0.770E-6 &  212300 & 0.004 & 0.2 &  85.4 & 0.005 & 0.64E+7 \\
19 & Telesto$^*$  & S &  12.3 & 0.260E-3 &  295000 & 0.001 & 1.2 &  36.6 & 0.021 & 0.13E+6 \\
20 & Calypso$^*$   & S &   9.5 & 0.120E-3 &  295000 & 0.001 & 1.5 &  59.1 & 0.026 & 0.32E+6 \\
21 & Helene     & S &  18. & 0.480E-3 &  377600 & 0.007 & 0.2 &  10.7 & 0.008 & 0.11E+6 \\
22 & Polydeuces$^*$ & S &   1.3 & 0.310E-6 &  377600 & 0.019 & 0.2 & 280.4 & 0.019 & 0.61E+9 \\
23 & Hyperion   & S & 135. & 0.370E+0 & 1481500 & 0.105 & 0.6 &   7.2 & 0.106 & 0.91E+6 \\
24 & Cordelia$^{\dagger}$  & U &  21. & 0.220E-2 &   49800 & 3E-4$^{\rm J98}$ & 0.2 &   2.6 & 0.004 & 0.20E+2 \\
25 & Ophelia$^{\dagger}$  & U &  23. & 0.290E-2 &   53800 & 0.011 & 0.1 &   7.3 & 0.011 & 0.95E+2 \\
26 & Bianca$^{\dagger}$   & U &  27. & 0.470E-2 &   59200 & 0.001 & 0.1 &   1.1 & 0.002 & 0.17E+2 \\
27 & Cressida$^{\dagger}$  & U &  41. & 0.170E-1 &   61800 & 6E-4$^{\rm F}$ & 0.1 &   0.6 & 0.002 & 0.10E+2 \\
28 & Desdemona$^{\dagger}$ & U &  35. & 0.100E-1 &   62700 & 7E-4$^{\rm F}$ & 0.1 &   0.8 & 0.002 & 0.13E+2 \\
29 & Juliet$^{\dagger}$    & U &  53. & 0.360E-1 &   64400 & 0.001 & 0.06$^{\rm F}$ &   0.3 & 0.001 & 0.81E+1 \\
30 & Portia$^{\dagger}$    & U &  70. & 0.820E-1 &   66100 & 5E-4$^{\rm F}$ & 0.02$^{\rm F}$ &   0.1 & 0.001 & 0.65E+1 \\
31 & Rosalind$^{\dagger}$  & U &  36. & 0.110E-1 &   69900 & 9E-4$^{\rm F}$ & 0.05$^{\rm F}$ &   0.5 & 0.001 & 0.17E+2 \\
32 & Cupid$^{\dagger}$    & U &   9. & 0.180E-3 &   74400 & 0.005 & 0.1 &   7.4 & 0.005 & 0.15E+4 \\
33 & Belinda$^{\dagger}$   & U &  45. & 0.220E-1 &   75300 & 8E-4$^{\rm F}$ & 2E-3$^{\rm F}$ &   0.2 & 0.001 & 0.17E+2 \\
34 & Perdita$^{\dagger}$   & U &  13.3 & 0.570E-3 &   76400 & 0.002 & 0.06$^{\rm F}$ &   2.1 & 0.002 & 0.18E+3 \\
35 & Puck$^{\dagger}$      & U &  81. & 0.130E+0 &   86000 & 1E-4$^{\rm F}$ & 0.3 &   0.8 & 0.005 & 0.20E+2 \\
36 & Mab$^{\dagger}$      & U &  12.4 & 0.460E-3 &   97700 & 0.003 & 0.1 &   3.1 & 0.003 & 0.93E+3 \\
37 & Miranda    & U & 235.8 & 0.430E+1 &  129900 & 0.001 & 4.4 &   2.7 & 0.077 & 0.96E+2 \\
38 & Naiad      & N &  29. & 0.853E-2 &   48200 & 1E-4$^{\rm B}$ & 4.7$^{\rm B}$ &  40.3 & 0.082 & 0.18E+3 \\
39 & Thalassa   & N &  40. & 0.236E-1 &   50100 & 2E-4$^{\rm B}$ & 0.2$^{\rm B}$ &   1.2 & 0.003 & 0.61E+1 \\
40 & Despina    & N &  74. & 0.117E+0 &   52500 & 3E-4$^{\rm B}$ & 0.04$^{\rm B}$ &   0.2 & 0.001 & 0.23E+1 \\
41 & Galatea    & N &  79. & 0.190E+0 &   62000 & 2E-4$^{\rm B}$ & 0.01$^{\rm B}$ &   0.0 & 0.000 & 0.48E+1 \\
42 & Larissa    & N &  96. & 0.255E+0 &   73500 & 0.001$^{\rm B}$ & 0.2$^{\rm B}$ &   0.5 & 0.004 & 0.75E+1 \\
43 & Hippocamp  & N &  17.4 & 0.150E-2 &  105300 & 1E-5$^{\rm B}$ & 2E-3$^{\rm B}$ &   0.0 & 0.000 & 0.16E+3 \\
44 & Proteus    & N & 208. & 0.258E+1 &  117600 & 5E-4$^{\rm B}$ & 0.04$^{\rm B}$ &   0.0 & 0.001 & 0.20E+2 \\
\enddata
\end{deluxetable*}

Fig. \ref{sesq} adds an important new dimension to the issue of sesquinary catastrophe. A moon may be vulnerable to sesquinary catastrophe in principle, but for an actual collisional cascade to happen, multiple ejecta reaccretion timescales must elapse since the moon acquired its current orbit. This explains why Pallene survives on its present orbit despite not being protected by a resonance. Our rough estimate of the reimpact time for Pallene is longer than 1~Myr. If Pallene escaped a resonance with Mimas relatively recently (on Myr timescales), it is to be expected that there has not been enough time for Pallene to undergo a collisional cascade. A faint, continuous ring associated with Pallene has been observed \citep{hed09}, possibly indicating a beginning of a collisional cascade. 

Having established $q_v$ and $\tau$ as the most relevant parameters for the sesquinary catastrophe we can now revisit the question whether it is relevant to populations other than small inner moons. For Ceres $q_v \approx 6$ and $\tau \approx 10^{12}$~yr, making ejecta reimpacts highly unlikely (other asteroids have both larger $q_v$ and $\tau$). For Saturn's largest irregular Phoebe $q_v \approx 3$ and $\tau \approx 5\times 10^9$~yr, making ejecta reaccretion possible but inefficient (both parameters have larger values for smaller irregulars). For Dimorphos, the satellite of km-sized near-Earth asteroid Didymos and the target of DART mission \citep{che23}, $q_v \approx 0.1$ and $\tau \approx 3$~days, making reccretion and/or ejection very efficient. Note that $\tau$ for Dimorphos presented here uses Eq. \ref{eq:tau} and is not applicable to real-life ejecta from Dimorphos, for which the inherited eccentricity is dwarfed by that arising from the ejection. These rough calculations confirm the reasoning presented in the Introduction that the sesquinary catastrophe is unlikely to be important for Solar System bodies other than the small inner moons of the giant planets and Mars.

In the next few sections we will address in more detail three cases of moons with $q_v>10$ that appear to be vulnerable to the sesquinary catastrophe: small resonant moons of Saturn, Neptune's Naiad which is also in a resonance, and Jupiter's Thebe which is not resonant.

\section{Small resonant Moons of Saturn}

All of the resonant moons of Saturn smaller than Epimetheus have the parameter $q_v>10$. This includes moons like Anthe and Methone that are in resonances with Mimas, as well as all of the Trojan moons in the system that orbit 60$^{\circ}$ ahead or behind larger moons. As these moons are relatively small and have small collisional cross sections, reimpact times of sesquinary ejecta are on the order of million years or longer, explaining why Pallene may have avoided sesquinary catastrophe in the time since it presumably left a resonance with a larger moon (possibly Mimas). But what about the moons that are still in the resonance, how does the resonance affect their ejecta?

As we mentioned in the previous section, resonances can both excite satellite orbits and plausibly help contain the ejecta. In the case of small resonant moons of Saturn which are not particularly eccentric and inclined (Fig. \ref{amd}), we think that the latter effect may be more important. However, not all resonances are equal and we need to consider different effects of Lindblad and coorotation resonances.

Lindblad resonances are those in which the eccentricity of the perturbed body is affected by the resonance, and a non-zero eccentrity of the perturbee is necessary to maintain this kind of resonance. A Lindblad resonance requires the resonant argument to contain the longitude of pericenter $\varpi$ of the perturbed body. A good example is the 4:3 Titan-Hyperion mean-motion resonance with the resonant argument $4 \lambda_H - 3 \lambda_T - \varpi_H$ ($\lambda$ stands for mean longitude, and subscripts T and H refer to Titan and Hyperion). Depending on the situation, perturbed bodies that are caught in the same kind of Lindblad resonances with the same perturber may still collide with each other at relatively high velocities (especially if they have different eccentricities). We will discuss similar resonances in the next section.

Corotation resonances, in contrast, depend on the eccentricity of the perturbing body and therefore include the perturber's $\varpi$ in the resonant argument. A corotation resonance can be stable even if the perturbee does not have any eccentricity of its own. An example is Anthe, which is in 10:11 resonance with Mimas \citep{coo08} with a resonant argument $11 \lambda_A - 10 \lambda_M - \varpi_M$ (where subscripts A and M refer to Anthe and Mimas). While there are multiple resonant islands associated with this corotation resonance, material that is in the same island with Anthe and has low proper ("free", as opposed to "forced") eccentricity and inclination should impact Anthe at very low velocities. This is also true of Methone which is in 14:15 resonance with Mimas. 

Anthe and Methone are associated with arcs of material, while non-resonant Pallene is associated with a full ring \citep{hed09}. This is clearly due to resonant perturbations from Mimas which confine longitudinally the material coming off the resonant moons \citep{hed09}. Furthermore, images of Methone taken by the Cassini spacecraft indicate that it is very close to being a smooth ellipsoid, possibly assembled from loose material \citep{tho13}. 

We conclude that the satellites Methone and Anthe are likely to retain some of the ejecta due to dynamical properties of their resonances. It is even possible that these bodies were assembled in the resonance, or are able to grow in the resonance due to accetion of material onto pre-existing cores \citep{tho13}. As long as both the moons and ejecta have low free eccentricity, re-impacts will happen at low velocities, even if the nominal eccentricity forced by Mimas is larger. Note that this may not be the complete explanation for the ongoing survival of the moons resonant with Mimas, as even when only the free eccentricity and inclination are taken into account, their $q_V$ factors are still little above 10. Still, we suspect that, in the case of Mimas's resonant companions, dynamical factors preclude the operation of the sesquinary catastrophe as we envision it. Unlike in the case of Jovian and Neptunian moons (next section) considerations based on both Roche limit and visual appearance of the moons suggest that small Saturnian moons may be rubble piles, and that (yet to be fully understood) resonant effects preclude the sesquinary catastrophe, rather than the moons' potentially higher material strength. We also speculate Pallene was likely in a similar resonance until recently, and that it may experience sesquinary catastrophe in the next few Myr. 

Trojan points are other potential locations for accumulation of loose material. However, Trojan moons that were imaged in good resolution (Telesto, Calypso, and Helene) do not appear to be ellipsoidal like Methone but have more irregular shapes, even if they appear blanketed by loose material \citep{tho13}. Behavior of their ejecta may be rather complex, with slowly ejected particles staying in tadpole orbits, while some faster fragments may be accreted by their parent moons \citep{dob10, nay16b, fer20, fer22a, fer22b}. Conversely, ejecta from the larger moons may impact the Trojans, and finally the Trojans may also exchange ejecta between themselves. In any case, sesquinary cratering of the Trojan moons is not well described by our parameter $q_v$ which assumes isolated moons, and more work is needed to determine if the sesquinary catastrophe is a possible outcome for these satellites.

The smaller horseshoe coorbital Epimetheus has a $q_v$ parameter of about 5 (based on its mean eccentricity), and the coorbitals are associated with a dusty ring \citep{por06}. Epimetheus has an irregular shape \citep{bur19, tho20} and is clearly not assembled on its present orbit from dust, especially as horseshoe orbits do not have stationary points like corotation resonances. We interpret the presence of the dusty ring as an indication that Epimetheus may experience some mass loss when being impacted by ejecta from Janus or/and its own ejecta pertubed by Janus, possibly making re-accretion of dust a slow and inefficient process. While Epimetheus is well short of our $q_v<10$ approximate condition for stability of satellites against sesquinary catastrophe, the presence of Janus likely complicates the mass loss from and accretion onto Epimetheus.

While macroscopic fragments largely follow gravitational trajectories, micron-sized dust is affected by radiation pressure and may acquire large eccentricities due to radiative perturbations \citep{bur79}. This may lead to reimpact velocities of these small particles to be much larger than their launch velocities, potentially leading to a self-sustaining process as proposed for E-ring particles emanating from Enceladus \citep{ham94}. While this additional mechanism may need to be considered when modelling dusty rings, it is unlikely to be important for large-scale collisional cascades we consider here, as larger fragments (which would constrain bulk of the mass) would not be as affected by radiation pressure as micron-sized dust.

\section{Naiad and Thebe}

Naiad is in a forth-order resonance with a more massive nearby moon Thalassa \citep{bro20}, so the behavior of its ejecta may plausibly be affected by this resonance. However, the Naiad-Thalassa resonance has the resonant argument $73 \lambda_T - 69 \lambda_N - 4 \Omega_N$ ($\Omega$ is the longitude of ascending node and subscripts N and T refer to Naiad and Thalassa), which indicates it affects only the inclination of Naiad and not that of Thalassa. This resonance is therefore a fourth-order inclination equivalent of a Lindblad resonance and does not have fixed points set by Thalassa that would be relevant to different bodies sharing the resonance with Naiad. Therefore it is highly unlikely Naiad could accrete from the material within the resonance due to Thalassa's perturbations. It is more likely that this resonance was a product of past convergent evolution of Thalassa and Naiad. 


Even if resonance were to offer some protection against high-velocity collisions between resonant bodies, it is unlikely Naiad's debris would stay in the resonance. Examination of the resonant width based on the librations measured by \citet{bro20} indicates the resonant width (in terms on Naiad's semimajor axis) on the order of 0.1~km. This is extremely narrow, but not surprising, given the small mass of Thalassa and the fourth order of the resonance. Given that the escape velocity of Naiad is about 25~m~s$^{-1}$, so the ejection velocity of the same order would place a fragment on an orbit roughly 100~km from that of Naiad, three orders of magnitude too far to stay in the resonance. Therefore, resonant confinement of debris is clearly not the explanation for the resilience of Naiad against sesquinary catastrophe.

A sesquinary catastrophe is not the only threat to the integrity of close-in inner moons, as some of them are very close to the Roche limit and experience very strong tidal forces. \citet{hed15} had already noted that some of the innermost satellites of Jupiter (Adrastea and Metis) and Neptune (Naiad, Thalassa and Despina) are within Roche limits expected for bodies held together only by gravity \citep{tis13}. While \citet{hed15} based their conclusion on the expected density of Neptunian inner moons, direct determination of Thalassa and Naiad's mass through resonant interaction  \citep{bro20} has confirmed that these satellites have low densities. In particular, Naiad's density of $\rho=800 \pm .48$~kg~m$^{-3}$ \citep{bro20} is well below the effective Roche density of 1700~kg~m$^{-3}$ at its distance from the planet \citep{tis13, hed15}. While the inner moons of Neptune appear to require internal strength, this is not true of the moons of Saturn and Uranus, where the distribution of rings and moons indicates that the satellites are under-dense rubble piles \citep{tis13, hed15}. Therefore we conclude that Naiad being an outlier in Fig. \ref{sesq} is probably best explained by Naiad (and possibly other innermost moons of Neptune) having substantial strength which changes the impact velocity for which partial accretion gives way to erosion. This is similar to what has been found for asteroids that are progenitors of asteroid families, which appear to have strength in excess of that expected from bodies held together only by gravity \citep{ste12}.

Thebe is another outlier that may exhibit material strength, even if it is less separated from (suspected) rubble piles in Fig. \ref{sesq} than Naiad. Thebe, unlike Naiad, is known to be associated with a dusty Thebe Gossamer Ring that probably consists of particles ejected off Thebe \citep{ock99, ham08}, indicating some erosion. However, Thebe Gossamer ring is the least dense of all the components of Jupiter's rings, so there is no indication that Thebe is losing material at a much higher rate than other three inner moons (Adrastea, Amalthea, and Metis). Amalthea is the only one among these moons for which density has been determined directly, and we know it is relatively under-dense at $\rho=860 \pm 100$ kg m$^{-3}$. \citet{hed15} point out that Metis and Adrastea need internal strength in order to stay intact against Jupiter's tidal forces, {\it if} they have the same density as Amalthea. Therefore we are left with two possible explanations of Thebe's continued existence on its excited orbit: either Thebe, Metis and Adrastea are unexpectedly more dense than Amalthea, or all of these moons have some internal strength. Either way the condition of $q_v > 10$ may not be sufficient to cause sesquinary catastrophe for these moons, and a higher threshold (e.g. $q_v=20$) is required.

While the inner moons of Neptune (and possibly Jupiter) appear to require internal strength, this is not true of the moons of Saturn and Uranus, where the distribution of rings and moons indicates that the satellites are under-dense rubble piles \citep{tis13, hed15}, which is consistent with our findings. In the case of Mars, while Phobos is within Roche limit by some criteria, it is not below the Roche density as found by \citet{tis13}:
\begin{equation}
\rho_R={4 \pi M_p  \over \gamma (a / R_p)^3 }
\label{roche}
\end{equation}
Assuming $\gamma=1.6$, the Roche density for Phobos is 1470~kg~m$m^{-3}$, as opposed to the actual value $\rho_P=1876$~kg~m$^{-3}$. Therefore it is possible (but not guaranteed) that Martian moons are similar  to those of Saturn and Uranus in having low strength. It is therefore possible Deimos may be relatively close to the stability limit for sesquinary catastrophe.

\section{Conclusions}

In his paper we recognize a new mechanism that can lead to destruction and re-accretion of small close-in planetary satellites, which we term sesquinary catastrophe. The basic idea is that any ejecta lost to planetocentric orbit from a satellite with a large eccentricity and/or inclination is likely to have a re-impact velocity that is much in excess of its ejection velocity. Large re-impact velocities would result from the fragment's orbit inheriting large eccentricity and/or inclination from the parent moon and then precessing out of alignment with the moon's orbit. If the collisions are fast enough to lead to net erosion of the moon, this will lead to a runaway process that may end with the destruction and likely re-accretion of the moon. 

To measure the susceptibility of the moon to sesquinary catastrophe, we define the parameter $q_v=\sqrt{e^2+\sin^2{i}}(v_{orb} / v_{esc})$. Theoretical arguments imply that $q_v$ of about 10 may be necessary for sesquinary catastrophe in rubble piles \citep{ste12}. We find that the small inner moons of Saturn that are likely rubble piles \citep{hed15} tend to have $q_v < 10$, unless their debris is shepherded by resonances, or the timescale for sesquinary catastrophe is very long. Saturnian moons which are close to being unstable against sesquinary catastrophe are also typically associated with dusty rings. The inner moons of Uranus, as well as those of Mars, also have $q_v<10$, consistent with them being rubble piles, too. We find that Neptune's highly-inclined moon Naiad must have internal strength, which has already been suggested on the basis of its vulnerability to tidal disruption \citep{hed15}. Similarly, Jupiter's Thebe needs to either have at least some strength, or must be significantly denser than Amalthea.

Probably the most important implication for satellite evolution is that sesquinary catastrophe makes it unlikely that small, close-in satellites can spend Gyrs on highly eccentric orbits. This would make the hypothesis of Phobos and Deimos being collisional fragments of one body \citep{bag21}, already criticized by \citet{hyo22}, even less plausible. Additional implication for the history of the Martian system is that past Martian moons that have reached significant eccentricties and inclinations \citep[as suggested by ][]{ros16} may have re-accreted on planar circular orbits through sesquinary catastrophe rather than having their orbits circularized by tides. This is particularly important for Deimos, which has a very low eccentricity that has been challenging to reconcile with the system's likely dynamically excited past \citep{yod82, bra20, qui20}.

In the Saturnian system, the sesquinary catastrophe could be an alternative mechanism of triggering satellite disruption and re-accretion, in addition to dynamical instability. However, sesquinary catastrophe is unlikely to affect the mid-sized satellites of Saturn (i.e. Mimas and larger), both because of their larger escape velocities and their more efficient tidal dissipation which may limit eccentricity growth. In the Uranian system the inner satellites are very closely packed, and are likely to experience dynamical instability \citep{dun97, fre12, cuk22} before their orbits are eccentric and/or inclined enough to be at risk of sesquinary catastrophe. In principle, sesquinary catastrophe could also affect the inner satellites of Jupiter and Neptune, as we show above the threshold for the effect may be higher in those systems because of higher internal strength of the moons.

In future work we hope to use direct numerical simulations that will clarify the potential for sesquinary catastrophe. Both the orbital evolution and outcomes of collisions will need to be simulated realistically to full confirm that this phenomenon does happen.

\begin{acknowledgments}
M\'C was supported by NASA Emerging Worlds Award 80NSSC19K0512. We thank the reviewer for their valuable comments.
\end{acknowledgments}

\bibliography{sesq.bib}{}
\bibliographystyle{aasjournal}



\end{document}